**Physarum Chip: Developments in growing computers from slime mould**

James G.H. Whiting [1], Ben P.J. de Lacy Costello [1,2], Andrew Adamatzky [1].

[1] Unconventional Computing Centre, University of the West of England, Bristol, UK.
[2] Institute of Biosensing Technology, University of the West of England, Bristol, UK.

The Phychip project is a collaborative European research initiative to design and implement computation using the organism *Physarum polycephalum*; it is funded by the Seventh Framework Programme (FP7) by the European Commission within CORDIS and the FET Proactive scheme. Included in this abstract are details the development of a Physarum based biosensor and biological logic gate, offering significant advancements in the respective fields.

*Physarum polycephalum* is a large single amoeba cell visible without a microscope, growing up to several centimetres across. This yellow slime mould forages for foodstuffs, extending protoplasmic tubes by a process of shuttle streaming. The cell is capable of optimising the shape of its protoplasmic networks in spatial configurations of attractants and repellents, resulting in spanning-tree arrangements. Such adaptive behaviour can be interpreted as computation. When exposed to attractants and repellents, Physarum changes patterns of its electrical activity associated with shuttle streaming, creating a peristaltic like action within the tubes, forcing cytoplasm along the lumen; the frequency of this oscillation determines the speed and direction of growth.

We experimentally derived a unique one-to-one mapping between a range of selected bioactive chemicals and the resultant oscillation voltage of the slime mould's extracellular electrical potential [1]. Amplitude and frequency change of this fluctuation can be used to identify specific chemicals with acceptable reliability; demonstrating detection of these chemicals in a similar manner to a biological contactless chemical sensor. The organism took approximately 10 minutes to respond fully to chemical exposure; in addition, 10 minutes of voltage recording was necessary to deduce frequency change. We believe results could be used in future designs of slime mould based chemical sensors and computers.

Surface electrical potential recordings were also made of a protoplasmic tube of the slime mould in response to a multitude of stimuli with regards to sensory fusion or multisensory integration and also to evaluate the effect of environmental fluctuations on the reliability of chemical detection [2]. Each stimulus was tested both alone and in combination to evaluate the effect that multiple stimuli have on the frequency of streaming oscillation. White light caused a decrease in frequency, whilst increasing the temperature and applying a food source in the form of oat flakes both increased the frequency. Simultaneously stimulating *P. polycephalum* with light and oat flake produced no net change in frequency, while combined light and heat stimuli showed an increase in frequency smaller than that observed for heat alone. When the two positive stimuli, oat flakes and heat, were combined, there was an overall increase in frequency similar to the cumulative increases caused by the individual stimuli. It is not known if Physarum polycephalum is able to differentiate between these different stimuli, but it is believed that the stimuli are detected by specific receptors and the effects are mediated through a multiple input, single output biochemical pathway resulting in sensory integration with amplitude and frequency of shuttle streaming. It is plausible that the calcium fluctuations responsible for shuttle streaming are the result of competitive or inhibitory binding of a bio-oscillator by receptor intermediaries.

The chemical biosensor does exhibit sensitivity to significant environmental changes such as increased light levels and large changes in temperature which may interfere with the voltage fluctuation and hence accuracy of analyte detection. It is proposed that this novel biosensor is capable of detecting many organic chemicals beyond those presented in this work and that the biosensor may be used for environmental monitoring and toxicity evaluation. The ability of Physarum to create sclerotia which can be reanimated years later offers significant advantages of shelf life in the biosensor field.

We also demonstrate that using these environmental and chemical stimuli we can use Physarum to approximate logic gates and derive combinational logic circuits [2], [3]. Changes in frequency as a result of non-equivalent inputs can be categorised into logic 1 or 0 depending on the desired gate. We have developed NOT, OR and AND gates, as well as the derived NOR, NAND, XOR and XNOR gates. For the NOT gate, light was used as an input; for the remaining gates, an increase in temperature was used for input A and an oat flake was used for input B. The gates are based around a single protoplasmic tube and are essentially programmable as the gate type can be chosen during operation without any hardware or wetware changes; the development of NAND or NOR gates means that the PFGs are functionally complete. Processing time for a single gate was approximately 20 to 30 minutes. The basic logic gates were reasonably accurate at predicting the correct output with AND/NAND giving the correct logic output 90% of the time while OR/NOR was 78% accurate. Multi gate combinational logic circuits such as XOR and Half Adder gave 71% and 65% accuracy respectively. Accuracy of the combinational logic decreases as the number of gates is increased, however they are at least as accurate as previous logic approximations using spatial growth of *Physarum polycephalum* and up to 30 times as fast at computing the logical output.

We have identified the analogue signal processing properties of a Physarum protoplasmic tube. Applying voltage waveforms to a tube and measuring the output we have observed that the tube acts like a first order low pass filter with a frequency cut-off of between 5kHz and 10kHz. The tube also acts to integrate square and sine waveforms at high frequencies; this has the potential to be used for analogue cell-based computing and hybrid organism electronics.

These results demonstrate the first steps towards Physarum computation and practical Physarum Biosensor; subsequent work will focus on development of a hybrid electronic-Physarum device capable of implementing computation.